\documentclass[aps,prl,preprint,groupedaddress,amsfonts,amssymb,amsmath]{revtex4}

\usepackage{bm}
\begin{document}

\title{A Topological String: The Rasetti-Regge Lagrangian, Topological
Quantum Field Theory, and Vortices in Quantum Fluids} 

\author{A.~D.~Speliotopoulos}

\email{adspelio@uclink.berkeley.edu}

\affiliation{National Research Council, Board on Physics and Astronomy,
2101 Constitution Avenue, NW, Washington, DC 20418}
\altaffiliation{Present address: Department of Physics, University of California at Berkeley, Berkeley, CA 94720-7300.}

\date{July 29, 2002}

\begin{abstract}
The kinetic part of the Rasetti-Regge action $I_{RR}$ for vortex lines
is studied and links to string theory are made. It is shown that both
$I_{RR}$ and the Polyakov string action $I_{Pol}$ can be constructed
with the same field $X^\mu$. Unlike $I_{NG}$, however, $I_{RR}$
describes a Schwarz-type topological quantum field theory. Using
generators of classical Lie algebras, $I_{RR}$ is generalized to
higher dimensions. In all dimensions, the momentum 1-form
$\mathbf{P}$ constructed from the canonical momentum for the vortex
belongs to the first cohomology class $H^1(M,\mathbb{R}^m)$ of the
worldsheet $M$ swept-out by the vortex line. The dynamics of the
vortex line thus depend directly on the topology of $M$. For a vortex
ring, the equations of motion reduce to the Serret-Frenet equations in
$\mathbb{R}^3$, and in higher dimensions they reduce to the
Maurer-Cartan equations for $so(m)$.   

\end{abstract}

\pacs{03.70.+k, 11.10.-z, 11.25.-w, 67.40.-w, 67.40.Vs}

\maketitle

\section{Introduction}
That vortex lines play an important role in many physical systems is
well-known \cite{Don, NO}. Following the original ideas of Onsager
\cite{Onsager} and Feynman \cite{Feynman}, researchers have even
used vortex rings in an attempt to explain the underlying cause of the
the lambda transition for superfluid He$^4$ \cite{Williams, LRU}.  
More recently, the discovery of Bose-Einstein condensates (BEC)
\cite{CW, W} has renewed interest in the study of vortex lines, and
within the last year vortex excitations have been seen experimentally
in BEC  \cite{MAHHWC, MCWD, CMD, AHWC}. There has been a corresponding
theoretical interest in the formation and stability of vortex lines in
BEC (see, for example, \cite{Rokshar} and \cite{GMPet}).

While experimental studies of vortex lines in quantum fluids have been
remarkable, theoretical understanding of the dynamics and interactions
of vortex lines on a quantum level has proceeded at a much slower
pace. Recent theoretical work on vortices in BEC has mostly been
focused only on the \textit{formation} and \textit{stability} of vortices in
the condensate and not on the properties and dynamics of the vortices
once they are formed. 

Much of the efforts in developing a deeper understanding of the
dynamics of vortex lines on a quantum level have been based on the
work of M.~Rasetti and T.~Regge. Using arguments from classical fluid
dynamics of ideal fluids, they \cite{RR1} proposed a lagrangian for
studying the quantum theory of vortex lines in quantum fluids in three
dimensions. Vortex lines are treated as extended objects and like a
string in string theory, a vortex line sweeps out a two-dimensional
worldsheet $M$ as it propagates in time. Current algebra methods
leading to the study of $Sdiff(\mathbb{R}^3)$, the diffeomorphic group
in $\mathbb{R}^3$, were then used by them and subsequent researchers
\cite{RR2, GMS1, PS1,AM1, AM2, GMS2, GM, PS2} in an effort to quantize
the field. 

In this paper we propose a different approach to understanding the
dynamics of vortex lines. Focusing on a single vortex, we start with
the field $X^\mu$ and make use of the $so(3)$ Lie algebra to rewrite
the kinetic part of the full Rasetti-Regge action $I_{RR}$ in terms of
differential forms, and demonstrate how $I_{RR}$ is related to the
Polyakov form \cite{Pol} of the Nambu-Goto action $I_{Pol}$
\cite{Nambu, Goto} (see also \cite{LR} for a  different approach). It
is then straightforward to see that while $I_{Pol}$ defines a
propagating string, $I_{RR}$ defines, when quantized, a Schwarz-type
topological quantum field theory (TQFT) \cite{Witten1, Witten2,
Sch}. Indeed, $I_{RR}$ is very similar in form to the Chern-Simons
lagrangian. Making use of other classical Lie algebras, we then extend
this construction to higher dimensions; the linkage between $I_{Pol}$
and $I_{RR}$ still hold (see also \cite{Ricca}). However, unlike
$I_{Pol}$, which can be constructed in any dimension, $I_{RR}$ exists
only in a discrete number of dimension corresponding to the dimensions
of the Lie algebra used in its construction.

Using this approach, it becomes clear that the understanding of the
quantum-and thus statistical behavior-of vortex lines will be the
first real world application of TQFT.  Conversely, the vortex system
provides a means of studying experimentally a TQFT for the first
time. The purpose of this paper is thus to make the connection
between $I_{RR}$ and TQFT, and our approach is strictly
classical, and our analysis formal. Nonetheless, a great deal can
immediately be discerned about the properties of vortex lines from the
fact that it is a TQFT and this classical analysis.

As is well known, a TQFT does not define a dynamical system in a
traditional sense; a single vortex line does not, strictly speaking,
have dynamical variables that evolve with time. TQFT's are interesting
nonetheless, \cite{Witten1, Witten2}. While our approach is strictly
classical, even at this level we find deep connections between topology
and the dynamics of vortices. Indeed, we show that the momentum 1-form
$\mathbf{P}$ constructed from the canonical momentum of the vortex
line belongs to the first cohomology class of $M$; the dynamics of
vortices depend directly on the topology of $M$. Going further, we
show formally that the solution to the equations of motion in
three-dimensions reduces to the Serret-Frenet equations for arbitrary
curves in $\mathbb{R}^3$. These equations are themselves equivalent to
the equation of motion of a charged particle constrained to  move on a
unit sphere in the the presence of a dyon located at the center. In
higher dimensions the equations of motion reduce to the Maurer-Cartan
equations for $so(m)$. The Maurer-Cartan 1-forms can be interpreted as
a ``pure gauge'' non-abelian vector potential, and as is the case for
TQFT, we are working with flat vector bundles. With this analogy,
explicit solutions of the equations of motion can be found using
Wilson path ordering. 

Connections between $I_{RR}$ and string theory goes beyond the
construction of $I_{Pol}$, however. A term of the form $\int
B_{\mu\nu}dX^\mu\wedge dX^\nu$, where $B_{\mu\nu}$ is a antisymmetric
tensor functional of the string field, was added to $I_{pol}$ by
Callan, Friedan, Martinec, and Perry \cite{CFMP} in their background
field treatment of string. $B_{\mu\nu}$ generates an
all-pervasive magnetic field in spacetime. While similar in form to
$I_{RR}$, in their treatment the specific functional dependence of
$B_{\mu\nu}$ on $X^\mu$ was determined by requiring that the trace anomally
of the \textit{total} string action vanish. This resulted in a
$B_{\mu\nu}$ that is dramatically different from what is considered
here. Along similar lines, Giveon, Rabinovici, and Veneziano 
\cite{GRV} also considered a $B_{\mu\nu}$ term in the string
lagrangian, but relaxed the trace anomally condition and considered
the effect of \textit{constant} $B_{\mu\nu}$ on the string.  

\section{General Construction of $I_{RR}$}

We begin with a classical, real Lie algebra $\mathfrak{g}$ with
generators $\mathbf{T}_a$ such that $[\mathbf{T}_a, \mathbf{T}_b] =
f_{ab}^{\>\>\>\>\> c} \mathbf{T}_c$, where $f_{ab}^{\>\>\>\>\> c}$ are
the structure constants for $\mathfrak{g}$, and indices run from
$1$ to $m$, the dimension of the $\mathfrak{g}$. $\mathbf{T}^a$ is
represented by matrices and following the convention in
\cite{LieAlg}, the Killing form $h_{ab} \equiv {\textrm
Tr\>} \{\mathbf{T}_a\mathbf{T}_b\} = -\delta_{ab}$ is
used to raise and lower indices: $A_a = h_{ab}A^b = -
A^b$. With this orthonormality condition, we can use the set
$\{\mathbf{T}_a\}$ as a natural basis for $\mathbb{R}^m$,
with $\mathbf{V}\in \mathbb{R}^m$ given by $\mathbf{V}
=  V^a \mathbf{T}_a$. The inner product on $\mathbb{R}^m$ is then $\langle
\mathbf{V}, \mathbf{U}\rangle \equiv -{\textrm Tr\>}
\{\mathbf{V}\mathbf{U}\}$ for $\mathbf{V}, \mathbf{U} \in 
\mathbb{R}^m$. Furthermore, using the identity matrix $\mathbf{I}$ of
$\mathfrak{g}$ we can extend this construction to the $m+1$-dimensional
Minkowski space $Min$ by taking $V = V_0\mathbf{I}/\sqrt{m} +
\mathbf{V}$ for $V\in Min$. When ${\mathfrak{g}} = su(2)$, this 
is just a representation of Minkowski space by the quarternions. 

We next consider an extended object $X^\mu(x^0, x^1)$ sweeping-out a
2-d surface $M$ in $Min$ where $x^0$ and $x^1$ are the spatial and time
coordinates on $M$. Taking $X = X_0
\mathbf{I}/\sqrt{m}+X_a\mathbf{T}^a$, the usual Nambu-Goto string 
action is obtained through 
\begin{eqnarray}
I_{Pol} &\equiv&-\textrm{Tr}\>\int dX\wedge *\>dX \nonumber \\
	&=& \int\sqrt{-g}g^{AB}\eta_{\mu\nu} \partial_A X^\mu
	\partial_B X^\nu dx^0 dx^1,  
\label{e2}
\end{eqnarray}
where capital roman indices run from $0$ to $1$, $d = dx^A \partial_A$
is the exterior derivative on $M$, $*$ is the Hodge $*$-operator, and
$g_{AB}$ is the  worldsheet metric. In this case $X^\mu$ describes a
string. 

The kinetic part of the Rasetti-Regge action is also constructed from
$X$, but now 
\begin{eqnarray}
I_{RR} &\equiv&-\frac{1}{3}\textrm{Tr}\>\int X dX\wedge dX 
       =-\frac{1}{3}\textrm{Tr}\> \int \mathbf{X}d\mathbf{X}\wedge d\mathbf{X} 
\nonumber \\
	&=&- \frac{1}{3}\int f_{abc}X^a \partial_0 X^b
	\partial_1 X^c dx^0 dx^1.
\label{e3}
\end{eqnarray}
$X^\mu$ in this case describes a vortex line. Note, however, that
$g_{AB}$ does not explicitly appear; $I_{RR}$ is a topological
invariant and describes a Schwarz-type TQFT similar to Chern-Simons
theory. (This corresponds to an antisymmetric-field langrangian in
background-field string theory with $B_{ab}\sim \epsilon_{abc}X^c$ in
three dimensions.) Note also that $I_{RR}$ is translationally
invariant; the lagrangian changes by a total derivative, $XdX\wedge
dX\to XdX\wedge dX+\Xi\, d(XdX)$  under the uniform translation $X \to
X + \Xi$. Indeed, $I_{RR}$ is the \textit{only} translationally
invariant topological action that can be constructed directly from
$X^\mu$. In the special case of $\mathfrak{g} = so(3)$, $I_{RR}$ is
proportional to the lagrangian in \cite{RR1}, but without the coupling
due to self-interaction. 

Notice that $I_{RR}$ does not depend on $X^0$, the time component of
$X^\mu$. This is expected: topological lagrangians describe systems
with no dynamical degrees of freedom. We will thus work solely with
$\mathbf{X}$ from this point on. This $\mathbf{X}$ is a 
section of a the vector bundle $\mathbb{R}^m$ over $M$, and is at
the same time an element of $\mathfrak{g}$, a \textit{vector} on
$\mathbb{R}^m$, and a 0-form (and thus a function) on $M$. Seen thusly,
the structure constants $f_{abc}$ form a rank-3, totally antisymmetric
tensor on $\mathbb{R}^m$. The 1-form $\mathbf{F} = \mathbf{F}_A dx^A =
F_{A}^a dx^A \mathbf{T}_a$ is then a \textit{vector valued} or,
equivalently, a Lie algebra valued 1-form on $M$, meaning that each of
its two \textit{components} $\mathbf{F}_A$ are both vectors in
$\textbf{R}^m$ and members of $\mathfrak{g}$.   

The equations of motion, $d\mathbf{X}\wedge d\mathbf{X}=0$, from
eq.~$(\ref{e3})$ can be integrated once to give 
\begin{equation}
\mathbf{P} \equiv [\mathbf{X}, d\mathbf{X}],
\label{e5}
\end{equation}
where $\mathbf{P}$ is a \textit{closed} Lie algebra valued 1-form on the
worldsheet: $d\mathbf{P} = 0$. The two components of $\mathbf{P}$ are
$\mathbf{P}_0\equiv P^c_0 \mathbf{T}_c 
= f_{ab}^{\>\>\>\>\> c}X^a\partial_0 X^b\mathbf{T}_c$ and
$\mathbf{P}_1\equiv P^c_1 \mathbf{T}_c= f_{ab}^{\>\>\>\>\>
c}X^a\partial_1 X^b \mathbf{T}_c$. $\mathbf{P}$ is related to the
canonical momentum for $\mathbf{X}$ through the dual form $\bm{\Pi}
\equiv *\> \mathbf{P}$:
\begin{equation}
\Pi^A_a = \frac{1}{\sqrt{-g}}\frac{\delta I_{RR}}{\delta\partial_A
 X^a},
\end{equation}
where $\bm{\Pi} = \Pi^a_A dx^A\mathbf{T}_a$. The components of
$\mathbf{P}$ then determine the momentum of the vortex, and we call
$\mathbf{P}$ the \textit{momentum 1-form}.

This choice for $\mathbf{P}$ is only unique up to a total
derivative. Although we could have just as well chosen
$\mathbf{P}'=2\mathbf{X}d\mathbf{X}$, $\mathbf{P}-\mathbf{P}' =
d\mathbf{X}^2$, and the two choices differ by an exact form. Indeed,
under uniform translations, $\mathbf{X} \to \mathbf{X} + \mathbf{K}$,
$\mathbf{P} \to  \mathbf{P} + d[\mathbf{K},\mathbf{X}]$, and
$\mathbf{P}$ changes by an \textit{exact} 1-form. Conversely, suppose
we have $\mathbf{P}_1 \equiv [\mathbf{X}_1, d\mathbf{X}_1]$ and
$\mathbf{P}_2 \equiv [\mathbf{X}_2, d\mathbf{X}_2]$ that differ by a
close form $d\mathbf{F}$. Then $d\mathbf{F} =
d[\mathbf{X}_2-\mathbf{X}_1, \mathbf{X}_2+\mathbf{X}_1]/2 - 
[\mathbf{X}_2+\mathbf{X}_1,d(\mathbf{X}_2-\mathbf{X}_1)]$ so that  either
$\mathbf{X}_2-\mathbf{X}_1 = \mathbf{K}$, or
$\mathbf{X}_2+\mathbf{X}_1 = \mathbf{K}$, where $\mathbf{K}$ is a
constant. Thus, $\mathbf{X}_2$ is related to $\mathbf{X}_1$ by either 
a uniform translation or a reflection plus a translation. Therefore,
what is physically relevant are the \textit{equivalence classes} of 
$\mathbf{P}$, where $\mathbf{P}_1 \sim \mathbf{P}_2$ if they differ 
by an exact form, and not any one specific choice of
$\mathbf{P}$. Consequently, $\mathbf{P}\in H^1(M,\mathbb{R}^m)$, the
first cohomology class of $M$, and  we are interested in $\mathbf{P}$
that are closed but \textit{not} exact.   

The cohomology classes for 2-d surfaces are well known \cite{Bott}. In
particular, $H^1(M,\mathbb{R})^m=0$ if $M$ is \textit{not} a closed
surface. This result has definite implications for the dynamics of
vortex lines: The dynamics of an open vortex line, which sweeps-out a
2-d open sheet in $\mathbb{R}^m$, differ dramatically from that of a
closed vortex line (vortex ring), which sweeps out a closed
surface. 

For the open vortex line, $H^1(M,\mathbb{R}^m)=0$ and we can
always make a translation to a frame in which the momentum vanishes,
$\mathbf{P} = 0$ so that $0=[\mathbf{X}, d\mathbf{X}]$. The
solution for $\mathbf{X}$ in this case is particularly simple. For
${\mathfrak{g}} = so(3), su(2), sp(2)$, $\mathbf{X} = a(x^0,x^1)
\mathbf{H}$, where $a$ is an arbitrary function and
$\mathbf{H}$ is a constant vector. The vortex line is constrained to
move along one direction: $\mathbf{H}$. For other Lie algebras,
$\mathbf{X} \in \mathfrak{c}$, the Cartan subalgebra for
$\mathfrak{g}$, so that $\mathbf{X} = X^i\mathbf{H}^i$ where   
$\{\mathbf{H}^i\}$ form the bases for $\mathfrak{c}$ \cite{LieAlg}, and
$\mathbf{X}$ propagates within a linear subspace of
$\mathbb{R}^m$.  

For the vortex ring, on the other hand, $H^1(M,\mathbb{R}^m) =
{\mathbb{Z}^m}$, the integers, and $\mathbf{P}$ need not
vanish. The dynamics of vortex rings are thus much more interesting,
and we shall focus on them for the rest of the paper. We begin with
$\mathfrak{g} = so(3)$, $su(2)$ or $sp(2)$. The vortex is propagating
in $\mathbb{R}^3$ and it's dynamics are especially contrained.

\section{Vortex Rings in $\mathbb{R}^3$}

When ${\mathfrak{g}} = so(3)$, $f_{abc} = \epsilon_{abc}$, and
we are dealing with a vortex line propagating in $\mathbb{R}^3$. Although
we can revert to the usual vector notation in this case, doing so will
add notational complexity. Instead, we introduce a slight abuse of
notation and write the cross product of two vectors $\mathbf{V},
\mathbf{U}\in \mathbb{R}^3$ as $\mathbf{V} {\textbf x}\mathbf{U}\equiv
[\mathbf{V}, \mathbf{U}]$.  

It is straightforward to show that $[\mathbf{P}_0,
\mathbf{P}_1] =0$; the two vectors are proportional to one
another. Consequently, we can write $\mathbf{P} = \mathbf{\hat{b}} p$
where $p$ is a scalar 1-form on $M$ and $\mathbf{\hat{b}}\in
\mathbb{R}^3$. (The hat denotes a unit vector:
$\lvert\mathbf{\hat{b}}\rvert^2=\langle\mathbf{\hat{b}},
\mathbf{\hat{b}}\rangle = 1$.)  Because $d\mathbf{P} = 0$,
$d\mathbf{\hat{b}} \wedge p + \mathbf{\hat{b}} dp = 0$; each term must
vanish separately. Consequently, $dp=0$, and $p$ is a closed 1-form. For the
other term, $d\mathbf{\hat{b}} \wedge p =0$, 
and from Cartan's lemma \cite{Spivak}, $d\mathbf{\hat{b}}$ must be
proportional to $p$: $d\mathbf{\hat{b}} = - \tau \mathbf{\hat{n}}p$,
where $\tau$ is an arbitrary function and $\mathbf{\hat{n}}$ is a unit
vector in $\mathbb{R}^3$ orthogonal to $\mathbf{\hat{b}}$. Doing this
trick once again and noting that $dd\mathbf{\hat{b}} = 0$,
$d\mathbf{\hat{n}} = (-\kappa \mathbf{\hat{t}} +\tau
\mathbf{\hat{b}})p$ where $\kappa$ is another arbitrary function on
$M$ and $\mathbf{\hat{t}} = \mathbf{\hat{n}}{\textbf x}
\mathbf{\hat{b}}$. Once again $\langle  \mathbf{\hat{n}},
\mathbf{\hat{t}}\rangle = 0$. It is then straightforward to show that
$d\mathbf{\hat{t}} = \kappa \mathbf{\hat{n}}$, and no more terms need
to be introduced.  

To complete the solution for $\mathbf{X}$, we note that
$\mathbf{\hat{t}}$, $\mathbf{\hat{n}}$, $\mathbf{\hat{b}}$ form a
moving orthogonal coordinate system on $\mathbb{R}^3$. Taking $\mathbf{X} =
\lvert\mathbf{X}\rvert(\alpha\mathbf{\hat{t}}+\beta\mathbf{\hat{n}}+
\gamma\mathbf{\hat{b}})$ for constants $\alpha, \beta,\gamma$, we
require that this $\mathbf{X}$ solves eq.~$(\ref{e5})$. Then $\alpha
=1$ and $\mathbf{X}$ lies along $\mathbf{\hat{t}}$, while
$\lvert\mathbf{X}\rvert^2 = 1/\kappa$. Solution of equations of motion
therefore reduces to finding solutions for
$\mathbf{\hat{t}}$, $\mathbf{\hat{n}}$, and $\mathbf{\hat{b}}$ for
given $\kappa$, $\tau$ and $p$.  

From $\langle \mathbf{\hat{b}}, \mathbf{\hat{n}}\rangle = 0$ and
$d\mathbf{\hat{t}} = \kappa \mathbf{\hat{n}}$, we see that $d\tau$ and
$d\kappa$ are both proportional to the 1-form $p$; the functions
$\kappa$, $\tau$  that determine $\mathbf{\hat{t}}$,
$\mathbf{\hat{n}}$, $\mathbf{\hat{b}}$ all depend upon
$p$. Consequently, there is a function $s(x^0,x^1)$ such that
\textit{locally} $ds = p$ and
\begin{eqnarray}
\mathbf{\hat{t}}' &=& \kappa(s) \mathbf{\hat{n}}, \nonumber \\
\mathbf{\hat{n}}' &=& -\kappa(s) \mathbf{\hat{t}}+\tau(s)
			\mathbf{\hat{b}}, \nonumber \\   
\mathbf{\hat{b}}' &=& -\tau(s) \mathbf{\hat{n}}, 
\label{e8}
\end{eqnarray}
where the prime denotes derivative wrt to $s$. These are the
Serret-Frenet equations \cite{Spivak} for a curve 
$\mathbf{c}: [a,b] \to \mathbb{R}^3$ parameterized by its arclength
$s$. $\kappa=1/\lvert{\mathbf X}\rvert^2$ is the local curvature of
$\mathbf c$ and is positive definite, as required, while $\tau$ is the
local torsion. Because $\mathbf P$ is a closed 1-form that is
\textit{not} exact, $\mathbf c$ is a \textit{closed} loop in
$\mathbb{R}^3$ \cite{Spivak}. 

The existence of solutions to the Serret-Frenet equations is guaranteed
\cite{Spivak}. It is nevertheless instructive to look further into
their explicit form for two special cases. Let $\kappa = \varpi
\cos u$, $\tau = \varpi \sin u$ where $u=u(s)$ and $-\pi/2\le u\le
\pi/2$ because $\kappa \ge 0$. Working with the coordinates $dt
=\varpi ds$, eqs.~$(\ref{e8})$ can be combined into
$\mathbf{\ddot{\hat{n}}} = -\mathbf{\hat{n}} + 
\dot{u}\>\mathbf{\hat{n}}{\mathbf{x}}\mathbf{\dot{\hat{n}}}$, where the
dot denotes derivative wrt $t$. This is similar to the equation of motion for
a particle constrained to move on a sphere in the presence of a
electric and magnetic dipole (a dyon) at the center of it, but
in this case the ratio of the magnetic to electric ``charge'' of the
dyon is $\dot{u}$ and can depend on time. Taking $\mathbf{\hat{l}} = 
\mathbf{\hat{n}}\mathbf{x}\mathbf{\dot{\hat{n}}}$, the torque
$\mathbf{\dot{\hat{l}}} =-\dot u \mathbf{\dot{\hat{n}}}$ is opposite of the
velocity of the particle $\mathbf{\dot{\hat{n}}}$ and has strength $\dot
u$. Consequently the total volume
$\mathbf{\hat{n}}\cdot\mathbf{\dot{\hat{n}}}\mathbf{x}
\mathbf{\ddot{\hat{n}}}$ swept out by $\mathbf{\hat{n}}$ is just
$\dot u$. If this volume is a constant, then taking $\omega =
\sqrt{1+\dot{u}^2}$, 
\begin{eqnarray}
\mathbf{\hat{t}} &=& 
	      \left\{
			\cos u
				\sin\left(\omega t\right)
			-
              		\frac{\dot u}{\omega}
	    		\sin u
				\cos\left(\omega t\right)
	      \right\}\mathbf{T}_1-\nonumber \\
	     &{}&
	      \left\{
			\cos u
				\cos\left(\omega t\right)
			+
              		\frac{\dot u}{\omega}
	      		\sin u
				\sin\left(\omega t\right)
	      \right\}\mathbf{T}_2 + 
		\frac{\sin{u}}{\omega}\mathbf{T}_3
	      \nonumber \\
\mathbf{\hat{b}} &=& 
	     - \left\{
			\sin u
				\sin\left(\omega t\right)
			+
              			\frac{\dot u}{\omega}
	    		\cos u
				\cos\left(\omega t\right)
		\right\}\mathbf{T}_1-\nonumber \\
	     &{}&
	      \left\{
			\sin u
				\cos\left(\omega t\right)
			-
              		\frac{\dot u}{\omega}
	      		\cos u
				\sin\left(\omega t\right)
			\right\}\mathbf{T}_2+
		\frac{\cos u}{\omega}\mathbf{T}_3
	      \nonumber \\
\mathbf{\hat{n}} &=& \frac{1}{\omega} 
	      \left\{
			\cos\left(\omega t\right)\mathbf{T}_1
			+
	      		\sin\left(\omega t\right)\mathbf{T}_2+
	      		\dot{u}\mathbf{T}_3
		\right\}.
\label{e10}
\end{eqnarray}
Furthermore, if $\dot{u}=0$, then $\mathbf{\hat{b}}= \mathbf{T}^3$ and
$\mathbf c$ is confined to the $1-2$ plane. Periodicity of
$\mathbf{\hat{t}}$ and $\mathbf{\hat{n}}$ for a closed curve
$\mathbf{c}$ gives $2\pi n =  \int_a^b \varpi ds = \int_a^b \varpi p$.
This is a well-known result \cite{Spivak} for closed curves and is the
fundamental reason why $p$ (and consequently $\mathbf P$) is a close
but not exact 1-form.  For general $u$, $0 =
\mathbf{\dddot{\hat{n}}}-\ddot{u}\mathbf{\ddot{\hat{n}}}/\dot{u}+(1+\dot{u}^2) 
\mathbf{\dot{\hat{n}}}-\ddot{u}\mathbf{\hat{n}}/\dot{u}$ with the
boundary conditions $\lvert \mathbf{\hat{n}}\rvert =1$,
$\lvert\mathbf{\dot{\hat{n}}}\rvert = 1$, and $\mathbf{\hat{n}}(0) = \mathbf{T}^1$.

\section{Vortex Rings in $\mathbb{R}^m$}

To solve eq.~$(\ref{e5})$ for general $\mathfrak{g}$ we follow an
approach similar to that in the previous section and
introduce a set of linearly independent vectors
$\{\mathbf{\hat{t}}_{r}\}\in \mathbb{R}^m$ on $M$ where
$\mathbf{\hat{t}}_r = R_r^{\>\>a}\mathbf{T}_a$ such that
$\langle\mathbf{\hat{t}}_r,\mathbf{\hat{t}}_s\rangle =
\delta_{r,s}$. The set $\{\mathbf{\hat{t}}_r\}$ forms a moving frame
on $\mathbb{R}^m$ for points on $M$ (letters in the second
have of the alphabet denote coordinates in the moving frame). Then
$R_{ra} R_{sa} = \delta_{rs}$ and $R\in so(m)$; similarly, $R_{ra}
R_{rb} = \delta_{ab}$. In addition, $[\mathbf{\hat{t}}_r,
\mathbf{\hat{t}}_s] = f_{rs}^{\>\>\>\>\> t} 
\mathbf{\hat{t}}_t$, but now $f_{rs}^{\>\>\>\>\>t}= R_r^{\>\>a}
R_s^{\>\>b}R^t_{\>\>c} f_{ab}^{\>\>\>\>\> c}$ are the ``structure
constants'' in the moving frame. Because they depend on $R_r^{\>\>a}$,
in this frame $f_{rs}^{\>\>\>\>\>t}$ need not be constant.

Since $\{\mathbf{\hat{t}}_r\}$ are orthonormal and span $\mathbb{R}^m$,
$d\mathbf{\hat{t}}_r = - \kappa_{rs}\mathbf{\hat{t}}_s$, where
$\kappa_{rs} = - \kappa_{sr}$ are 1-forms on $M$. Moreover,
from $dd\mathbf{\hat{t}}_r =0$,
\begin{equation}
0 = d\kappa_{rs}+\kappa_{rt}\wedge\kappa_{ts}.
\label{e12}
\end{equation}
These are the Maurer-Cartan equations \cite{Spivak} and $\kappa_{rs}$
are the Maurer-Cartan 1-forms for $so(m)$. Indeed, let $S^{\tilde a}$
be the generators of $so(m)$, the symmetry group of
$\mathbb{R}^m$, such that $[S^{\tilde a},S^{\tilde b}] = k^{\tilde
a\tilde b}_{\>\>\>\>\>\tilde c}S^{\tilde c}$, ${\rm Tr\>}\{S^{\tilde
a}S^{\tilde b}\}=-\delta^{\tilde a\tilde b}$, and $\tilde a$ runs from
$1$ to $m(m-1)/2$. For a fixed $\tilde a$, $S^{\tilde a}$ are $m\times
m$ antisymmetric matrices with elements $(S^{\tilde a})_{rs}$ (we are
\textit{not} working in the adjoint representation for $so(m)$). We
then introduce the 1-forms $A= A^{\tilde a}S^{\tilde a}$ with values
in the Lie algebra $so(m)$ such that $\kappa_{rs}\equiv (A^{\tilde
a}S^{\tilde a})_{rs}$. Then $dA+A\wedge A=0$; $A$ can be seen as a
non-abelian``vector potential'' for the group $so(m)$. The field
strength for $A$ vanishes, however, and $A$ is a ``pure gauge'' vector
potential. As expected, $A$ does not contain any physical degrees of
freedom. Indeed, written in terms of matrices of $\mathit{so(m)}$,
$R^{t}dR=-A$.

With this interpretation of the Maurier-Cartan equations it is
straightforward to see that
\begin{equation} 
\mathbf{\hat{t}}_r = {\textrm P} \Bigg(\exp{\int_0^s
A}\Bigg)_{r}^{\>\>\>\>\>a}\mathbf{T}_a,
\label{e13}
\end{equation}
where P denote Wilson path ordering.

Solution to eq.~$(\ref{e5})$ now follows straightforwardly. Given a set of
$\kappa_{rs}$, we construct $\mathbf{\hat{t}}_r$ using
eq.~$(\ref{e13})$. We then choose $\mathbf{X} = \lvert\mathbf{X}\rvert
\mathbf{\hat{t}}_1$ so that $\mathbf{P} = \lvert\mathbf{X}\rvert^2
f_{1rs}\kappa_{1r}\mathbf{\hat{t}}_s$. Because $d\mathbf{P}=0$,
\begin{equation}
0=\{d\log{\lvert\mathbf{X}\rvert^2}f_{1r's'}-\kappa_{1t'}f_{tr's'}\}
\wedge\kappa_{1r'},
\label{e14}
\end{equation}
where $r', s', t' >1$ and we have used $df_{rst} =
-\kappa_{rn}f_{nst}-\kappa_{sn}f_{rnt}-\kappa_{tn}f_{rsn}$.
In addition, the choice of $\kappa_{rs}$ must satisfy the constraint
$0=f_{1r's'}\kappa_{1r'}\wedge\kappa_{1s'}$; $\{\kappa_{1r'}\}$
therefore can \textit{not} linearly independent. One  
solution of this constraint equation is $\kappa_{1r'} = \kappa_{r'}
\pi$, where $\kappa_a$ are functions on $M$ and $\pi$ is a 1-form on
$M$. This choice of $\kappa_{1r'}$ does not restrict $\kappa_{r's'}$ and
$0=d\kappa_{r's'}+\kappa_{r't'}\wedge\kappa_{t's'}$ still. Integration
of eq.~$(\ref{e14})$ then gives $\lvert\mathbf{X}\rvert^2 =
\exp\{\int\alpha  \pi\}$ for any function $\alpha$ on $M$, and we are
done. $\mathbf{X}$ is determined by the arbitrary function $\alpha$,
and Maurer-Cartan 1-forms $\kappa_{r'}\pi$, and $\kappa_{r's'}$. 

Except for $so(3)$, $su(2)$, and $sp(2)$, this choice of $\kappa_{1r'}$ is
not the most general one that satisfies the constraint
equation. Indeed, with this choice, $\mathbf{P}_0\propto \mathbf{P}_1$ 
and like the $\mathbb{R}^3$ case, the two components of $\mathbf{P}$ are
proportional to one another. It is expected that when the general
solution to the constraint equation is used, this relationship between
the components of $\mathbf{P}$ will no longer hold. 

\section{Concluding Remarks}

We have shown in this paper the deep connection between 
$I_{RR}$, on the one hand, and string theory and TQFT on the
other. Indeed, the topological nature of the theory, and the
fundamental role it plays in determining vortex dynamics, is manifest
in our approach in analyzing the system. Moreover, with this approach
generalization of vortex dynamics to higher dimensions becomes
straightforward.

With the goal being to establish the links between TQFT,
string theory, and the study of vortex lines, the approach we have
taken in this paper has been purposefully formal. We have focused on
establishing mathematical structures, and using these structures in
understanding the general physical properties of vortices propagating in
superfluids that is due solely to the kinetic part of the full
Rasette-Regge lagrangian. 

We have focused on only the kinetic part of the lagrangian for two
reasons. First, the traditional interaction term between 
vortex lines found in \cite{RR1} is extremely 
nonlinear. Some degree of perturbative analysis, based on the
kinetic term, would most likely be needed. To this end, a thorough
understanding of the ``free'' kinetic term is needed. Second, the
interaction term has a $1/r$ type of divergence singularity at the
classical level, which has traditionally been regulated by introducing
a finite vortex core. However, it is expected that the degree of
divergence will be weakened in the full quantum field theoretic
treatment of the system, and a complete treatment of this divergence
will most fruitfully be delayed until then. The first step in the
quantum field theoretic approach is to the quantization of the
``free'' (kinetic) part of the Rasette-Regge lagrangian $I_{RR}$. 

How to treat the many-vortex system is still an open question. Once
more than one vortex line is introduced, a whole host of questions
come to the fore. One particular issue is the question of how
interactions between them should be incorporated into the approach
outlined here. One can certainly choose to use the classical
interaction term found in \cite{RR1}. Another approach could be
to follow the approach of string theory where the interaction of
strings are  represented by the merging and breaking of strings (which
for closed strings fundamentally changes the genus, and thus topology,
of the surface it sweeps-out). Much of the techniques developed for
string theory could then conceivably be applied to the analysis of
interacting vortex lines. Which of these two approaches will be more
fruitful is unclear, especially in light of the two points listed
above.    

The question of how to include interactions between vortices goes
beyond a discussion of field-theoretic techniques and methodology,
however. As we have mentioned in the introduction, a TQFT has no
dynamics in the traditional sense; since the lagrangian does not
depend on the metric, there is no notion of time. Will the inclusion
of the interaction terms necessitate the introduction of the metric?
While it is possible to use de Rham's method of generalized forms
\cite{deRham} to rewrite and generalize the interaction term found in
\cite{RR1} in terms of differential forms (which will thus
automatically be independent of the metric), it is unclear if such an  
approach is physically meaningful. Moreover, making sense of this
interaction term will require the introduction of a high energy
cut-off (the vortex core size), which may bring along its own 
particular set of problems. 




\end{document}